\def\year{2021}\relax
\documentclass[letterpaper]{article} % DO NOT CHANGE THIS
\usepackage{aaai22}  % DO NOT CHANGE THIS
\usepackage{times}  % DO NOT CHANGE THIS
\usepackage{helvet} % DO NOT CHANGE THIS
\usepackage{courier}  % DO NOT CHANGE THIS
\usepackage[hyphens]{url}  % DO NOT CHANGE THIS
\usepackage{graphicx} % DO NOT CHANGE THIS
\def\UrlFont{\rm}  % DO NOT CHANGE THIS
\usepackage{natbib}  % DO NOT CHANGE THIS AND DO NOT ADD ANY OPTIONS TO IT
\usepackage{caption} % DO NOT CHANGE THIS AND DO NOT ADD ANY OPTIONS TO IT
\usepackage{cite}
\usepackage{amsmath,amssymb,amsfonts}
\usepackage{algorithmic}
\usepackage{graphicx}
\usepackage{textcomp}
\usepackage{xcolor}
\usepackage{tabularx}
\usepackage{multirow}
\usepackage[linesnumbered,ruled,vlined]{algorithm2e}
\def\BibTeX{{\rm B\kern-.05em{\sc i\kern-.025em b}\kern-.08em
    T\kern-.1667em\lower.7ex\hbox{E}\kern-.125emX}}
\title{Cybersecurity Anomaly Detection in Adversarial Environments}
\author{
    David A. Bierbrauer\textsuperscript{\rm 1}, Will Kritzer\textsuperscript{\rm 2}, Alexander Chang\textsuperscript{\rm 3}, Nathaniel D. Bastian\textsuperscript{\rm 4}\\
}
\affiliations{
    Army Cyber Institute, United States Military Academy\textsuperscript{\rm 1,4}, Johns Hopkins University\textsuperscript{\rm 2,3}\\
    \textsuperscript{\rm 1}david.bierbrauer@westpoint.edu, \textsuperscript{\rm 2}wkritze1@jhu.edu, \textsuperscript{\rm 3}achang66@jhu.edu, \textsuperscript{\rm 4}nathaniel.bastian@westpoint.edu
}

\begin{document}

\maketitle

\begin{abstract}
The proliferation of interconnected battlefield information-sharing devices, known as the Internet of Battlefield Things (IoBT), introduced several security challenges. Inherent to the IoBT operating environment is the practice of adversarial machine learning, which attempts to circumvent machine learning models. This work examines the feasibility of cost-effective unsupervised learning and graph-based methods for anomaly detection in the network intrusion detection system setting, and also leverages an ensemble approach to supervised learning of the anomaly detection problem. We incorporate a realistic adversarial training mechanism when training supervised models to enable strong classification performance in adversarial environments. The results indicate that the unsupervised and graph-based methods were outperformed in detecting anomalies (malicious activity) by the supervised stacking ensemble method with two levels. This model consists of three different classifiers in the first level, followed by either a Naive Bayes or Decision Tree classifier for the second level. The model maintains an F1-score above 0.97 for malicious samples across all tested level two classifiers. Notably, Naive Bayes is the fastest level two classifier averaging 1.12 seconds while Decision Tree maintains the highest AUC score of 0.98. 
\end{abstract}

\section{Introduction}
In the information age, most human activity depends on the computer network to function, and our dependence on digital processes for productivity means that system failure would lead to a tremendous loss of resources and inefficiency. This is especially true with the proliferation of the Internet of Battlefield Things (IoBT). The concept of the IoBT relies on highly-connected devices to provide timely and accurate information across the battlefield. Unfortunately, advances in this technology present several unique security challenges \cite{Kamhoua2020}. Should these devices fall victim to an attack, malicious actors could disrupt mission critical information flows and inhibit a commander's ability to command and control across the full spectrum of operations. Historically, such attacks have been identified, recognized, and dealt with by cybersecurity professionals who used manual tools to scan the network traffic for suspicious activity. The emerging techniques of machine learning (ML) have the potential to make these professionals even more effective. If a ML  model could identify anomalous packets of data traveling through an IoBT network autonomously and accurately, then human security professionals would waste less time sifting through network traffic alerts and log files.

For such a ML model to be implemented, analysts would like low levels of false positives and negatives from the ML decision outputs. The ML model would also require efficiency so that organizations could run it on standard computing equipment. We set out to build such a model for anomaly detection in cybersecurity suitable for implementation according to the above criteria.

In developing a suitable model for anomaly detection, we must consider methods that account for the adversarial environment where adversaries could use adversarial machine learning (AML) techniques; these are categorized based on the adversary's knowledge of the target model. In a white-box AML method, the attacker knows the architecture or parameters of their target, and black-box techniques are used when they do not have such information \cite{Kose2019}. Some defensive schemes have been proposed to harden learning models, but new AML methods are constantly being developed.

Cybersecurity professionals need novel techniques to aid them in identifying malicious attacks while simultaneously maintaining a low rate of false positives and false negatives in adversarial IoBT environments. This work explores the feasibility of some unsupervised methods, as well as a graph-based approach to anomaly detection. We also develop a supervised stacking ensemble model trained on realistic adversarial samples that maintains a high level of precision and recall.

\section{Related Work}

\subsection{Common Approaches}

ML techniques for tackling the anomaly detection problem in cybersecurity have included semi-supervised methods, deep learning models, and graph-based approaches. A common baseline approach for semi-supervised anomaly detection is the One Class Support Vector Machine (OC-SVM) \cite{AHMED201619}, which finds the hyperplane that separates the data from the origin with the greatest possible distance, using the same kernel mechanics as a traditional SVM \cite{nguyen2018scalable}. OC-SVM has been highly effective on noisy data sets in cybersecurity, but it struggles with run-time on data sets with high dimensionality \cite{8233268}.

As network traffic data is generally high-dimensional, deep learning approaches have supplanted OC-SVM as the industry standard. A primary deep learning technique for anomaly detection in cybersecurity is the autoencoder (AE), where a neural network is tasked with reconstructing network traffic patterns, after having passed them through a series of ever-smaller hidden layers. Its central layer is a reduced-dimensions representation of normal traffic, since it is trained exclusively on normal traffic. When the network is fully trained, the instances that are still inaccurately reconstructed by the AE are considered to be anomalous \cite{7966342}. 

The most elaborate implementation of the AE in cybersecurity is the variational autoencoder (VAE), a directed graph-based probabilistic model that leans on an AE's central layer for learning model parameters \cite{An2015VariationalAB}. There is skepticism from researchers that AEs are not any more effective than the classic OC-SVM, but the academic consensus is that AEs can be a useful tool, especially in conjunction with more classic methods \cite{nguyen2018scalable}.

Two additional unsupervised ML methods for anomaly detection are relevant. Isolation forests exploit a decision tree-like architecture to uncover anomalies, assuming that those anomalies would be most easily split from the rest of the data (and thus closer to the root of the decision tree) \cite{8283275}. Local outlier factor is a common density-based unsupervised learning method, which identifies anomalies as points which have significantly lower density than neighboring observations \cite{XU20131174}.

Anomaly detection can also be performed by modeling computer networks as graphs. Although anomaly detection is a well-researched problem, the vast majority of the prior approaches have treated networks as static graphs. Static graph-based methods had severe limitations, as they failed to capture the temporal characteristics of emerging anomalies. Microcluster-based Detector of Anomalies in Edge Streams (MIDAS) is a novel anomaly detection technique, which uses dynamic graphs to detect micro-cluster anomalies in the network. MIDAS scans network traffic to find sudden groups of suspiciously similar edges in dynamic graphs. MIDAS has relatively robust predictive power, and its chief advantage is detecting anomalies quickly in real-time \cite{SB2020}.

\subsection{Adversarial Attacks}

Malicious actors will attempt to circumvent the network intrusion detection system (IDS) via adversarial attacks, often employing AML techniques. Although AML techniques attempt to exploit different attack vectors, they largely have the same goal: cause a model to misclassify samples. In the context of IoBT, an adversary may want to gain and maintain access to the network of devices. They may have several goals associated with that access, such as stealing information or degrading sensor capabilities. No matter the goal, the adversary will need to avoid detection by any ML-enabled IDS platforms. They can accomplish this in many ways, including poisoning data during the collection process to reduce the prediction quality of the model. Evasion attacks entail supplying a perturbed sample to a trained model with the goal of misclassifying that sample. In cybersecurity, this means an adversary would perturb malicious traffic (e.g., slightly modify features) to mask it as normal \cite{shipp2020advancing}. One such AML technique for evasion attacks, for example, is the Fast Gradient Sign Method (FGSM). With FGSM, adversarial examples are generated according to Equation \ref{eqn:FGSM}:
\begin{equation}
    x_{adv} = \epsilon\cdot\text{sign}(\nabla_{x}J(\theta,x,y)),
    \label{eqn:FGSM}
\end{equation}
where $\epsilon$ is a scale parameter, $J$ is the loss function, $\theta$ are the model parameters, $x$ are the inputs (features), and $y$ are the targets (labels) \cite{GSS2016}. Another effective technique is the Carlini-Wagner Attack, which is formulated as a minimization problem primarily to construct adversarial image samples \cite{CW2017}.

Defending against these adversarial evasion attacks has received widespread attention in recent years. One approach uses adversarial training leveraging an ensemble-based stacking method. In this method there are two levels: the first stack consists of multiple classifiers whose output is a new feature matrix of the predicted labels from each classifier; and the second stack is a single classifier that outputs a final prediction. The first stack classifiers are trained on an original training set, as well as some adversarial training sets. Classifiers in the first stack are chosen based on their performance against an unchanged test set \cite{DB2021}. This approach served as a primary motivator for our own ensemble method.

\section{Methodology}

We briefly describe the cybersecurity data set used and pre-processing performed, and give an overview of the unsupervised and graph-based methods explored for anomaly detection. We also describe the process for developing the adversarial training sets, and the subsequent ensemble approach for adversarial training based supervised learning.

\subsection{Data Set}

We used the UNSW-NB15 data set synthesized by the Australian Center for Cybersecurity \cite{MS2015, MS2016}. It contains 2.5 million observations of packets traveling through a computer network, where each observation is labelled as normal traffic or attack traffic. The set contains 321,283 malicious observations and 2.2 million normal observations. Many of the attacks were artificially injected into the real traffic to make the entire data set more representative of all the types of attacks an anomaly detection system could encounter.

UNSW-NB15 contains 47 features of which 42 are numerical and 5 are nominal. In addition to these 47 features, the data included two labels: the attack category for malicious traffic, and a binary label for normal (0) and malicious (1) traffic. These features described the characteristics of the packet flow, the quantity of information contained in each packet, and the broad features of the packet content, among other characteristics. There are an additional 12 pre-engineered features developed by the authors included as part of the 47 features provided in the data set.

\subsection{Pre-Processing}

We first examined the data set to check for unrealistic entries. For example, we found a few rows whose ``source port" value did not fall in the range of actual port values and, therefore, removed them from the data set. We also removed the data set author's pre-engineered features as we worked under the assumption that our ML model only had intrinsic network data with no pre-determined logic applied. We also encoded the nominal features as pre-processing for modeling. After exploring feature importance through a $\chi^2$ test and a Mutual Information test, we dropped several of the encoded features that were not in the top five results during importance testing. As such, we then examined a correlation matrix between remaining features, which allowed us to drop any that were highly-correlated (greater than 0.85). Through this process, we were able to reduce the dimensionality of the data set from 49 to 21 features. With this reduced data set, which was used for all experiments moving forward, we divided the data into training and test sets to evaluate our ML trained classifiers. The original, unchanged training set with 21 features is called $T_1$. It is important to note that the test set remains unchanged (i.e., it is not perturbed as part of the adversarial training) in order to provide a standard basis for ML model evaluation.

\subsection{Unsupervised Learning and Graph-based Methods}

We first experimented with unsupervised and graph-based methods for anomaly detection using the UNSW-NB15 data set. For the unsupervised learning, we implemented both the isolation forest and local outlier factor methods on $T_1$. An isolation forest selects a random feature and a random split value between that feature's maximum and minimum value. The algorithm continues and builds an isolation tree and anomalous samples are those with a smaller path length in the tree \cite{Liu2008}. The local outlier factor method, on the other hand, measures the deviation of the samples with respect to their neighbors. Samples with lower density than their neighbors are considered outliers \cite{BKNS2000}. These techniques helped us determine whether the anomalous cluster could be separated from the normal instances in an unassisted yet effective manner. 

We then analyzed the data with MIDAS, a graph-based approach to anomaly detection. To do this, we extracted the following features from the original UNSW-NB15 data set: timestamp, source IP address, and destination IP address. We organized the data set in ascending chronological order in terms of timestamp and ran the MIDAS algorithm. The MIDAS algorithm takes as input a stream of graph edges over time using the features described above (e.g., the source and destination IPs are the nodes, and a sample at time $t$ provides the edge). For efficiency, the state of the graph is stored in Count-Min-Sketch (CMS) data structures to keep count of the number of edges between nodes. There are two such CMS structures. First, we maintain a count $s_{uv}$, which is the total number of edges over time between nodes $u$ and $v$. The second is the number of edges $a_{uv}$ at the current time. The primary difference is that $s_{uv}$ is maintained while $a_{uv}$ is reset when we move forward to the next time tick. These structures can be queried for the approximate number of edges $\hat{s_{uv}}$ and $\hat{a_{uv}}$. While new edges between nodes $u$ and $v$ are provided to MIDAS, an anomaly score at time $t$ is output according to Equation \ref{eqn:MIDAS}:

\begin{equation}
    \text{score}((u,v,t)) = \bigg(\hat{a}_{uv}-\frac{\hat{s}_{uv}}{t}\bigg)^2\frac{t^2}{\hat{s}_{uv}(t-1)}.
    \label{eqn:MIDAS}
\end{equation}
This score is based on a $\chi^2$ goodness-of-fit test under the assumption that the mean rate at which edges appear at time $t$ is the same as the mean rate for all times before $t$. Further details of of this algorithm can be found in \citet{SB2020}.

Our goal in experimenting with these methods was strictly focused on the anomaly detection task in a non-adversarial environment. Likewise, we focused on non-adversarial anomaly detection for MIDAS since it only used specific features to assign anomaly scores.

\subsection{Development of Adversarial Training Sets}

In preparing to perform supervised learning for anomaly detection using the UNSW-NB15 data set, we sought to train models robust against adversarial evasion attacks at inference time as part of a network intrusion detection system. Thus, we used adversarial training as part of an ensemble approach \cite{DB2021}, which required the generation of adversarial training sets. As such, we generated these adversarial training sets according to realistic methods, assuming the adversary could influence (i.e., perturb) approximately 20\% of the network traffic. We employed two different approaches, which had an overarching goal of increasing the number of false negatives; that is, we wanted to confuse the classifiers so that malicious samples would be classified as normal. The first approach was inspired by FGSM. This method, which in our case considers a 0-1 loss function, uses a linear discriminant analysis (LDA) decision function to determine a direction in which each sample's features are perturbed. The goal is essentially to shift samples across the decision boundary to confuse the classifier. The algorithm used can be seen in Algorithm \ref{algo:FGSM}. The resulting training set is $T_2$.

\begin{algorithm}[hbt!]
	\DontPrintSemicolon % Some LaTeX compilers require you to use \dontprintsemicolon instead
	\KwIn{$n \times 1$ Training Label Vector $Y$, $m \times n$ Training Feature Matrix $X$, Adjustment Factor $\epsilon$}
	\KwOut{Adversarial Training Set $X_{adv}$}
	Fit LDA model using standardized $X$ and $Y$. Save the model coefficients in $1 \times m$ vector $W$\\
	Define a $n \times 1$ vector $V$\\
	\For{$k \in 1\leq k\leq n$}{
		Calculate Decision Function: $d_k=\log P(y_k=1 | x_k) - \log P(y_k=0 | x_k)$\\
		\uIf{$d_k > 0$}{$V_k=-1$}
		\Else{$V_k=1$}
	}
	Calculate direction matrix $Z$: $Z=\text{sign}(VW)$\\ 
	\Return{$X_{adv}=X+\epsilon Z$}\;
	\caption{LDA FGSM}
	\label{algo:FGSM}
\end{algorithm}

The next adversarial training set we developed rested on the concept that some features held more importance to our model. The importance of certain features was determined previously using the $\chi^2$ test and the Mutual Information test, and then we examined those for features that an adversary could realistically control. For example, a reasonably controlled feature would be setting the source time-to-live for a given connection or fixing the amount of bytes sent from a source during that connection. The idea would then be to perturb the malicious samples by a fixed amount such that the feature mean for the malicious samples would closely approximate that of the normal samples as seen in Algorithm \ref{algo:FeaturePerturbation}. This final training set is $T_3$.

\begin{algorithm}[hbt!]
	\DontPrintSemicolon % Some LaTeX compilers require you to use \dontprintsemicolon instead
	\KwIn{$m\times n$ Training Feature Matrix $X$, $n\times 1$ Training Label Vector $Y$, Selected Feature List $F$}
	\KwOut{Adversarial Training Set $X_{adv}$}
	$X_{adv}=X$\\
	Set $N_0$ as the number of samples where true label is 0\\
	Set $N_1$ as the number of samples where true label is 1\\
	\For{$k \in F$}{
		Set $s_0=0$ and $s_1=0$\\
		\For{$i\in 1\leq i\leq n$}{
			\uIf{$Y_i = 0$}{$s_0=s_0+X_{i}^{k}$}
			\Else{$s_1=s_1+X_{i}^{k}$}
		}
		$\Delta_x=s_0-s_1$\\
		\For{$i \in 1\leq i\leq n$}{
			\uIf{$Y_i = 1$}{$X_{adv,i}^{k} = X_{i}^{k}\frac{N_1}{N_0}+\frac{\Delta_x}{N_0}$}
			\Else{$X_{adv,i}^{k} = X_{i}^{k}$}
		}
	} 
	\Return{$X_{adv}$}\;
	\caption{Feature Importance Perturbation}
	\label{algo:FeaturePerturbation}
\end{algorithm}

With the training sets defined, we then standardized our training and test feature matrices. Since we wanted to build supervised ML models on the original and adversarial training sets, the fixed test set was standardized according to the training set used for each iteration.

\subsection{The Ensemble Approach for Supervised Learning}
\begin{figure}[hbt!]
    \centering
    \includegraphics[scale=.65]{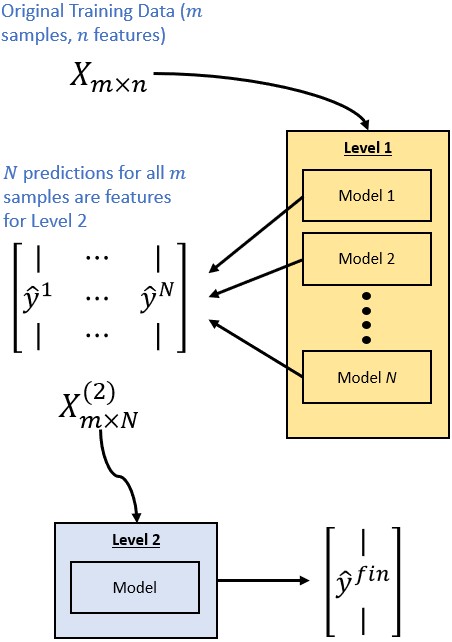}
    \caption{Stacking Ensemble Method Structure}
    \label{fig:ensemble}
\end{figure}

The ensemble approach for adversarial training based supervised learning for anomaly detection was inspired by the approach proposed in \citet{DB2021} which utilizes a two-level structure. This structure consists of multiple models in the first level that serve as base classifiers. These classifiers output either the class prediction (hard vote) or class probability (soft vote) as features for the next level. Level two then consists of a single classifier that learns from these new features to provide final class predictions. A visual representation of this ensemble approach is seen in Figure \ref{fig:ensemble}.

For the level one stack, we first experimented with several common classifiers trained with each data set. We then chose the best classifier for each data set ($T_1$, $T_2$, $T_3$) and optimized the classifiers' hyper-parameters. This optimization was accomplished by tuning according to the data set it performed best against. Next, we re-trained and passed each classifiers' class prediction probabilities as features to several candidate level two classifiers. In this way, we utilized a soft-voting stacking model proposed by \citet{DB2021}. Our level two stack then ideally consists of the single best classifier across all three data sets based on the new feature matrix. 

The primary metrics used to evaluate our classifiers were time-to-train and F1-score. We chose F1-score because of the imbalance between normal and malicious samples. Since our primary aim was successfully identifying malicious samples, we focused on the F1-score related to the malicious samples. We also considered the Area Under Curve (AUC) metric when results were inconclusive.

\section{Computational Experimentation}

Our experiments relied on scikit-learn, or $sklearn$, implementations in Python for feature-selection, pre-processing, and modeling \cite{scikit-learn}. The feature-selection module contains useful $\chi^2$ and mutual information tests, as each takes as input a feature matrix $X$ and label vector $y$. The $\chi^2$ test computes the $\chi^2$ statistic between each feature and class, while the mutual information test estimates dependency between variables.

For the unsupervised learning methods, the ensemble module of $sklearn$ provides an isolation forest implementation that provides an anomaly score for a feature matrix based on the extra tree regressor. The local outlier factor method is implemented through the neighbors module and also provides an anomaly score. The graph-based MIDAS approach, however, relied on a custom class developed by the authors \cite{bhatiaGitHub}.

The diverse supervised learning methods provided by $sklearn$ allow several classifiers to choose from. First, however, we generated the adversarial training sets using custom implementations of Algorithm \ref{algo:FGSM} and Algorithm \ref{algo:FeaturePerturbation}. We applied Standard Scaler from the $sklearn$ pre-processing package to normalize training and test set features before building classifiers. We used seven different classifiers drawn from different modules, including LDA, Quadratic Discriminant Analysis (QDA), Gaussian Naive Bayes, Bagging (with a Decision Tree base classifier), Decision Tree, Random Forest, and Logistic Regression. For Random Forest, we ensured we set the $n\_jobs$ parameter to use all available processing cores to increase performance. Once classifiers were selected for our level one stack, we optimized the classifier hyper-parameters using the grid search cross-validation method.

\section{Results and Discussion}

We will briefly describe the experimental results of the implemented unsupervised, graph-based, and supervised methods for anomaly detection using the UNSW-NB15 data set.

\subsection{Unsupervised Learning and Graph-based Methods}

As part of the computational experimentation, we tested both isolation forest and local outlier factor techniques as the baseline unsupervised methods. Unfortunately, the results were not on par with our expectations, as both methods achieved an AUC score of approximately 0.57 and a recall of less than 25\% on $T_1$. Less than a quarter of the relevant observations were correctly identified, which is not operable. With these baseline metrics, we experimented with the MIDAS technique, running the algorithm on the subset of the network traffic using the domain name system protocol. Due to MIDAS' model assumptions, we were unable to run it on observations using other protocols. Although MIDAS' final AUC score turned out to be 0.74, the algorithm only took 100 seconds to complete with approximately 800,000 rows of data. We concluded that graph-based MIDAS was a superior approach compared to the two unsupervised methods, given its extremely fast run-time and better AUC. 

\subsection{Stacking Ensemble Based Supervised Learning}

Subsequently, we implemented the stacking ensemble model using various supervised learning classifiers. In order to select our level one stack, we first trained and tested a variety of classifiers against our original and adversarial training sets. The results for F1-score are displayed in Table \ref{tab:level1F1}, where the F1-scores are reported separately for each class in the format of [F1-Normal, F1-Malicious].

\begin{table}[hbt!]
	\centering
	\begin{tabularx}{\columnwidth}{X|
	                                >{\centering\arraybackslash}X
	                                >{\centering\arraybackslash}X
	                                >{\centering\arraybackslash}X
	                                }
		\hline
		Classifier & $T_1$ & $T_2$ & $T_3$\\
		\hline
		\multirow{2}{5em}{LDA} & \multirow{2}{5em}{[.987, .915]} & \multirow{2}{5em}{[.987, .914]} & \multirow{2}{5em}{\textbf{[.987, .915]}}\\
		&&&\\
		\multirow{2}{5em}{QDA} & \multirow{2}{5em}{[.990, .934]} & \multirow{2}{5em}{[.990, .934]} & \multirow{2}{5em}{[.941, .719]}\\
		&&&\\
		\multirow{2}{5em}{Naive Bayes} & \multirow{2}{5em}{[.967, .817]} & \multirow{2}{5em}{[.981, .884]} & \multirow{2}{5em}{[.931, .688]}\\
		&&&\\
		\multirow{2}{5em}{Bagging} & \multirow{2}{5em}{[.990, .934]} & \multirow{2}{5em}{\textbf{[.991, .940]}} & \multirow{2}{5em}{[.990, .934]}\\
		&&&\\
		\multirow{2}{5em}{Decision Tree} & \multirow{2}{5em}{[.990, .934]} & \multirow{2}{5em}{\textbf{[.991, .940]}} & \multirow{2}{5em}{[.990, .934]}\\
		&&&\\
		\multirow{2}{5em}{Random Forest} & \multirow{2}{5em}{[.989, .927]} & \multirow{2}{5em}{[.989, .929]} & \multirow{2}{5em}{[.935, .704]}\\
		&&&\\
		\multirow{2}{5em}{Logistic Regression} & \multirow{2}{5em}{\textbf{[.991, .938]}} & \multirow{2}{5em}{[.990, .935]} & \multirow{2}{5em}{[.991, .936]}\\
		&&&\\
	\end{tabularx}
	\caption{Level 1 F1-Scores}
	\label{tab:level1F1}
\end{table}

From this initial experimentation, the following classifiers performed best for level one: Logistic Regression ($T_1$), Decision Tree ($T_2$), and LDA ($T_3$). We chose Decision Tree over Bagging for $T_2$ primarily due to its faster run-time. We chose LDA for $T_3$ to diversify our models so that we did not use the same model as that chosen for $T_2$. We then used the grid search method for hyper-parameter optimization and determined the optimal parameters based on F1-score for these methods. Next, we re-trained our level one stack consisting of these tuned classifiers and passed the new feature matrix to several classifiers to help us determine an optimal level two classifier. The results for F1-score and training time are depicted in Tables \ref{tab:level2F1} and \ref{tab:level2TrainTime}, respectively. Since we clearly did not have a consensus ``best" classifier among those tested for level two, we also considered the AUC metric; these results are shown in Table \ref{tab:level2AUC}.

\begin{table}[hbt!]
	\centering
	\begin{tabularx}{\columnwidth}{X|
	                                >{\centering\arraybackslash}X
	                                >{\centering\arraybackslash}X
	                                >{\centering\arraybackslash}X
	                                }
		\hline
		Classifier & $T_1$ & $T_2$ & $T_3$\\
		\hline
		\multirow{2}{5em}{LDA} & \multirow{2}{5em}{[.99728165, .98121153]} & \multirow{2}{5em}{[.99715777, .98032807]} & \multirow{2}{5em}{[.99691815, .978665]}\\
		&&&\\
		\multirow{2}{5em}{QDA} & \multirow{2}{5em}{\textbf{[.99728347, .98122352]}} & \multirow{2}{5em}{\textbf{[.99715867, .98033419]}} & \multirow{2}{5em}{[.99691905, .97867111]}\\
		&&&\\
		\multirow{2}{5em}{Naive Bayes} & \multirow{2}{5em}{\textbf{[.99728347, .98122352]}} & \multirow{2}{5em}{\textbf{[.99715867, .98033419]}} & \multirow{2}{5em}{[.99691905, .97867111]}\\
		&&&\\
		\multirow{2}{5em}{Bagging} & \multirow{2}{5em}{[.99728347, .98122329]} & \multirow{2}{5em}{[.99715688, .98032147]} & \multirow{2}{5em}{[.99691726, .9786581]}\\
		&&&\\
		\multirow{2}{5em}{Decision Tree} & \multirow{2}{5em}{[.99728165, .98121153]} & \multirow{2}{5em}{[.99715777, .98032807]} & \multirow{2}{5em}{[.99691815, .978665]}\\
		&&&\\
		\multirow{2}{5em}{Random Forest} & \multirow{2}{5em}{[.99728347, .98122352]} & \multirow{2}{5em}{\textbf{[.99715867, .98033419]}} & \multirow{2}{5em}{[.99691905, .97867111]}\\
		&&&\\
		\multirow{2}{5em}{Logistic Regression} & \multirow{2}{5em}{[.99728437, .98122964]} & \multirow{2}{5em}{\textbf{[.99715867, .98033419]}} & \multirow{2}{5em}{\textbf{[.99692086, .97868332]}}\\
		&&&\\
	\end{tabularx}
	\caption{Level 2 F1-Scores}
	\label{tab:level2F1}
\end{table}

\begin{table}[hbt!]
	\centering
	\begin{tabularx}{\columnwidth}{X|
	                                >{\centering\arraybackslash}X
	                                >{\centering\arraybackslash}X
	                                >{\centering\arraybackslash}X
	                                }
		\hline
		Classifier & $T_1$ & $T_2$ & $T_3$\\
		\hline
		\multirow{2}{5em}{LDA} & \multirow{2}{3em}{5.0369} & \multirow{2}{3em}{4.6606} & \multirow{2}{3em}{4.3494}\\
		&&&\\
		\multirow{2}{5em}{QDA} & \multirow{2}{3em}{2.2604} & \multirow{2}{3em}{1.9409} & \multirow{2}{3em}{1.8105}\\
		&&&\\
		\multirow{2}{5em}{Naive Bayes} & \multirow{2}{3em}{\textbf{1.3273}} & \multirow{2}{3em}{\textbf{1.0903}} & \multirow{2}{3em}{\textbf{0.9490}}\\
		&&&\\
		\multirow{2}{5em}{Bagging} & \multirow{2}{3em}{74.6318} & \multirow{2}{3em}{59.8093} & \multirow{2}{3em}{57.8239}\\
		&&&\\
		\multirow{2}{5em}{Decision Tree} & \multirow{2}{3em}{9.3937} & \multirow{2}{3em}{8.9551} & \multirow{2}{3em}{8.7126}\\
		&&&\\
		\multirow{2}{5em}{Random Forest} & \multirow{2}{3em}{33.1345} & \multirow{2}{3em}{32.7135} & \multirow{2}{3em}{37.9840}\\
		&&&\\
		\multirow{2}{5em}{Logistic Regression} & \multirow{2}{3em}{11.9229} & \multirow{2}{3em}{11.9820} & \multirow{2}{3em}{11.8420}\\
		&&&\\
	\end{tabularx}
	\caption{Level 2 Training Times}
	\label{tab:level2TrainTime}
\end{table}

\begin{table}[hbt!]
	\centering
	\begin{tabularx}{\columnwidth}{X|
	                                >{\centering\arraybackslash}X
	                                >{\centering\arraybackslash}X
	                                >{\centering\arraybackslash}X
	                                }
		\hline
		Classifier & $T_1$ & $T_2$ & $T_3$\\
		\hline
		\multirow{2}{5em}{LDA} & \multirow{2}{5em}{0.9894396} & \multirow{2}{5em}{0.9884303} & \multirow{2}{5em}{0.9874015}\\
		&&&\\
		\multirow{2}{5em}{QDA} & \multirow{2}{5em}{0.9894361} & \multirow{2}{5em}{0.9884312} & \multirow{2}{5em}{0.9874024}\\
		&&&\\
		\multirow{2}{5em}{Naive Bayes} & \multirow{2}{5em}{0.9894361} & \multirow{2}{5em}{0.9884312} & \multirow{2}{5em}{0.9874024}\\
		&&&\\
		\multirow{2}{5em}{Bagging} & \multirow{2}{5em}{0.9894597} & \multirow{2}{5em}{\textbf{0.9884451}} & \multirow{2}{5em}{\textbf{0.9874136}}\\
		&&&\\
		\multirow{2}{5em}{Decision Tree} & \multirow{2}{5em}{0.9894597} & \multirow{2}{5em}{\textbf{0.9884451}} & \multirow{2}{5em}{\textbf{0.9874136}}\\
		&&&\\
		\multirow{2}{5em}{Random Forest} & \multirow{2}{5em}{\textbf{0.9981971}} & \multirow{2}{5em}{0.9781675} & \multirow{2}{5em}{0.9804987}\\
		&&&\\
		\multirow{2}{5em}{Logistic Regression} & \multirow{2}{5em}{0.9824521} & \multirow{2}{5em}{0.9807217} & \multirow{2}{5em}{0.9791699}\\
		&&&\\
	\end{tabularx}
	\caption{Level 2 AUC Scores}
	\label{tab:level2AUC}
\end{table}

From these classification model performance metrics, there are several options for the level two classifier. Since all results are relatively good, we recommend using Naive Bayes if computational efficiency is a priority. Otherwise, we recommend using a Decision Tree here as its AUC score for $T_2$ and $T_3$ were highest and it maintained a high AUC for $T_1$; it is also a highly interpretable model.

\section{Conclusions}
Our results suggest that a stacking ensemble approach for supervised learning with LDA FGSM and feature importance perturbation methods used for adversarial training could be highly effective in detecting anomalies in the network intrusion detection setting, even if malicious actors are using AML to conduct evasion attacks against the model. With AUC scores of over 0.98 and total training times of less than one minute, our model could certainly be useful in IoBT settings, as long as they are able to collect data in a similar format to our data set.

There is also room for improving the ensemble approach. We did not attempt to optimize the hyper-parameters of the level two classifiers, so improvements could be made to the F1-score. Also, we did not attempt training the models on smaller amounts of data. It is possible that we will not sacrifice too much accuracy by training on a smaller number of observations, which could improve run-time greatly. Additionally, we could expand the stack to three levels. This would allow us to build out the level two stack with multiple classifiers that then pass prediction probabilities to a third level for a final classification. This approach would, of course, decrease overall performance in terms of run-time. Depending on the effectiveness of other adversarial methods, this trade-off may be worthwhile.

In future experimentation, we will explore other ways to test our methods. Since we did not consider adversarial training samples for our unsupervised approaches, future work should consider novel approaches for generating such samples. This would also allow for a more direct comparison between the unsupervised and supervised methods. The adversarial training methods we did incorporate, however, are certainly not the only approaches that could be used. We could also consider using other methods, such as adapting the Carlini-Wagner attack to our data set, or leverage other techniques from evolutionary computation and deep learning to generate adversarial examples as part of the adversarial training mechanism \cite{AMB2020}. This might reveal other strengths and weaknesses of classifiers within the stack. We also plan to incorporate adversarial examples into the training of the unsupervised learning methods, as recently done in \citet{HSU2021}.

Another recommendation for future work is to incorporate the MIDAS graph-based approach into the  ensemble model, perhaps leveraging other graph mining techniques such as graph neural networks. This would require adjusting the MIDAS algorithm to incorporate all types of network traffic and could be used to generate a new feature for classification purposes. We also should consider different types of malicious behavior rather than just a 0-1 classification model. In a real-world setting, professionals may prioritize different types of malicious activity to focus their efforts. This means we should consider a model that classifies these different attacks once identified as anomalous. Finally, we also believe conducting a test of our model through implementation on a closed network will validate our approach. We recommend using commonly available hardware - such as a Raspberry Pi - to examine model feasibility on cost-effective platforms, which would directly advise validity in the IoBT environment.

\section{Acknowledgements}
This work was supported by the U.S. Army Combat Capabilities Development Command (DEVCOM) Army Research Laboratory under Support Agreement No. USMA21050 and the U.S. Army DEVCOM C5ISR Center under Support Agreement No. USMA21056. The views expressed in this paper are those of the authors and do not reflect the official policy or position of the United States Military Academy, the United States Army, the United States Department of Defense, or the United States Government.

\bibliography{refs}

\begin{thebibliography}{22}
\providecommand{\natexlab}[1]{#1}

\bibitem[{Ahmed, {Naser Mahmood}, and Hu(2016)}]{AHMED201619}
Ahmed, M.; {Naser Mahmood}, A.; and Hu, J. 2016.
\newblock A survey of network anomaly detection techniques.
\newblock \emph{Journal of Network and Computer Applications}, 60: 19--31.

\bibitem[{Alhajjar, Maxwell, and Bastian(2021)}]{AMB2020}
Alhajjar, E.; Maxwell, P.; and Bastian, N. 2021.
\newblock Adversarial machine learning in Network Intrusion Detection Systems.
\newblock \emph{Expert Systems with Applications}, 186(115782): 1--13.

\bibitem[{An and Cho(2015)}]{An2015VariationalAB}
An, J.; and Cho, S. 2015.
\newblock Variational Autoencoder based Anomaly Detection using Reconstruction
  Probability.
\newblock Technical report, SNU Data Mining Center.

\bibitem[{Bhatia et~al.(2020)Bhatia, Hooi, Yoon, Shin, and Faloutsos}]{SB2020}
Bhatia, S.; Hooi, B.; Yoon, M.; Shin, K.; and Faloutsos, C. 2020.
\newblock MIDAS: Microcluster-Based Detector of Anomalies in Edge Streams.
\newblock \emph{Proceedings of the AAAI Conference on Artificial Intelligence},
  34(04): 3242--3249.

\bibitem[{Bhatia et~al.(2021)Bhatia, Liu, Hooi, Yoon, Shin, and
  Faloutsos}]{bhatiaGitHub}
Bhatia, S.; Liu, R.; Hooi, B.; Yoon, M.; Shin, K.; and Faloutsos, C. 2021.
\newblock MIDAS.
\newblock \url{https://github.com/Stream-AD/MIDAS}.

\bibitem[{Breunig et~al.(2000)Breunig, Kriegel, Ng, and Sander}]{BKNS2000}
Breunig, M.~M.; Kriegel, H.-P.; Ng, R.~T.; and Sander, J. 2000.
\newblock LOF: Identifying Density-Based Local Outliers.
\newblock \emph{SIGMOD Rec.}, 29(2): 93–104.

\bibitem[{Carlini and Wagner(2017)}]{CW2017}
Carlini, N.; and Wagner, D. 2017.
\newblock Towards Evaluating the Robustness of Neural Networks.
\newblock In \emph{2017 IEEE Symposium on Security and Privacy (SP)}, 39--57.

\bibitem[{Devine and Bastian(2021)}]{DB2021}
Devine, S.; and Bastian, N. 2021.
\newblock An Adversarial Training Based Machine Learning Approach to Malware
  Classification under Adversarial Conditions.
\newblock In \emph{Proceedings of the 54th Hawaii International Conference on
  System Sciences}, 827--836.

\bibitem[{Ghanem et~al.(2017)Ghanem, Aparicio-Navarro, Kyriakopoulos,
  Lambotharan, and Chambers}]{8233268}
Ghanem, K.; Aparicio-Navarro, F.~J.; Kyriakopoulos, K.~G.; Lambotharan, S.; and
  Chambers, J.~A. 2017.
\newblock Support Vector Machine for Network Intrusion and Cyber-Attack
  Detection.
\newblock In \emph{2017 Sensor Signal Processing for Defence Conference
  (SSPD)}, 1--5.

\bibitem[{Goodfellow, Shlens, and Szegedy(2015)}]{GSS2016}
Goodfellow, I.; Shlens, J.; and Szegedy, C. 2015.
\newblock Explaining And Harnessing Adversarial Examples.
\newblock In \emph{International Conference on Learning Representations}.

\bibitem[{Hsu et~al.(2021)Hsu, Chen, Lu, Lu, and Yu}]{HSU2021}
Hsu, C.; Chen, P.; Lu, S.; Lu, S.; and Yu, C. 2021.
\newblock Adversarial Examples for Unsupervised Machine Learning Models.
\newblock \emph{CoRR}, abs/2103.01895.

\bibitem[{Kamhoua et~al.(2020)Kamhoua, Njilla, Kott, and Shetty}]{Kamhoua2020}
Kamhoua, C.~A.; Njilla, L.~L.; Kott, A.; and Shetty, S. 2020.
\newblock \emph{Introduction}, chapter~1, 1--26.
\newblock John Wiley \& Sons, Ltd.
\newblock ISBN 9781119593386.

\bibitem[{Kose(2019)}]{Kose2019}
Kose, U. 2019.
\newblock Techniques for Adversarial Examples Threatening the Safety of
  Artificial Intelligence Based Systems.
\newblock \emph{CoRR}, abs/1910.06907.

\bibitem[{Liu, Ting, and Zhou(2008)}]{Liu2008}
Liu, F.~T.; Ting, K.~M.; and Zhou, Z.-H. 2008.
\newblock Isolation Forest.
\newblock In \emph{2008 Eighth IEEE International Conference on Data Mining},
  413--422.

\bibitem[{Moustafa and Slay(2015)}]{MS2015}
Moustafa, N.; and Slay, J. 2015.
\newblock UNSW-NB15: a comprehensive data set for network intrusion detection
  systems (UNSW-NB15 network data set).
\newblock In \emph{2015 Military Communications and Information Systems
  Conference (MilCIS)}, 1--6.

\bibitem[{Moustafa and Slay(2016)}]{MS2016}
Moustafa, N.; and Slay, J. 2016.
\newblock The Evaluation of Network Anomaly Detection Systems: Statistical
  Analysis of the UNSW-NB15 Data Set and the Comparison with the KDD99 Data
  Set.
\newblock \emph{Information Security Journal: A Global Perspective}, 25(1-3):
  18--31.

\bibitem[{Nguyen and Vien(2019)}]{nguyen2018scalable}
Nguyen, M.-N.; and Vien, N.~A. 2019.
\newblock Scalable and Interpretable One-Class SVMs with Deep Learning and
  Random Fourier Features.
\newblock In Berlingerio, M.; Bonchi, F.; G{\"a}rtner, T.; Hurley, N.; and
  Ifrim, G., eds., \emph{Machine Learning and Knowledge Discovery in
  Databases}, 157--172. Cham: Springer International Publishing.
\newblock ISBN 978-3-030-10925-7.

\bibitem[{Pedregosa et~al.(2011)Pedregosa, Varoquaux, Gramfort, Michel,
  Thirion, Grisel, Blondel, Prettenhofer, Weiss, Dubourg, Vanderplas, Passos,
  Cournapeau, Brucher, Perrot, and Duchesnay}]{scikit-learn}
Pedregosa, F.; Varoquaux, G.; Gramfort, A.; Michel, V.; Thirion, B.; Grisel,
  O.; Blondel, M.; Prettenhofer, P.; Weiss, R.; Dubourg, V.; Vanderplas, J.;
  Passos, A.; Cournapeau, D.; Brucher, M.; Perrot, M.; and Duchesnay, E. 2011.
\newblock Scikit-learn: Machine Learning in {P}ython.
\newblock \emph{Journal of Machine Learning Research}, 12: 2825--2830.

\bibitem[{Shipp et~al.(2020)Shipp, Clouse, Lucia, Ahiskali, Steverson, Mullin,
  and Bastian}]{shipp2020advancing}
Shipp, T.~J.; Clouse, D.~J.; Lucia, M. J.~D.; Ahiskali, M.~B.; Steverson, K.;
  Mullin, J.~M.; and Bastian, N.~D. 2020.
\newblock Advancing the Research and Development of Assured Artificial
  Intelligence and Machine Learning Capabilities.
\newblock In \emph{Proceedings of the AAAI Fall 2020 Symposium on AI in
  Government and Public Sector}. arXiv:2009.13250.

\bibitem[{Xu et~al.(2017)Xu, Wang, Meng, and Zhang}]{8283275}
Xu, D.; Wang, Y.; Meng, Y.; and Zhang, Z. 2017.
\newblock An Improved Data Anomaly Detection Method Based on Isolation Forest.
\newblock In \emph{2017 10th International Symposium on Computational
  Intelligence and Design (ISCID)}, volume~2, 287--291.

\bibitem[{Xu et~al.(2013)Xu, Yeh, Lee, and Li}]{XU20131174}
Xu, L.; Yeh, Y.-R.; Lee, Y.-J.; and Li, J. 2013.
\newblock A Hierarchical Framework Using Approximated Local Outlier Factor for
  Efficient Anomaly Detection.
\newblock \emph{Procedia Computer Science}, 19: 1174--1181.
\newblock The 4th International Conference on Ambient Systems, Networks and
  Technologies (ANT 2013), the 3rd International Conference on Sustainable
  Energy Information Technology (SEIT-2013).

\bibitem[{Yousefi-Azar et~al.(2017)Yousefi-Azar, Varadharajan, Hamey, and
  Tupakula}]{7966342}
Yousefi-Azar, M.; Varadharajan, V.; Hamey, L.; and Tupakula, U. 2017.
\newblock Autoencoder-based feature learning for cyber security applications.
\newblock In \emph{2017 International Joint Conference on Neural Networks
  (IJCNN)}, 3854--3861.

\end{thebibliography}
\end{document}